\documentclass[prl,aps,twocolumn,superscriptaddress,10pt,amsmath,amssymb,nofootinbib,floatfix]{revtex4-1}
\usepackage{epsfig}
\usepackage{amsmath}
\usepackage{amssymb}
\usepackage{graphicx}
\usepackage{enumitem}
\usepackage{longtable}

\makeatletter 

\begin{document}

\title{Effect of dopant ordering on the stability of ferroelectric
  hafnia}

\author{Sangita Dutta} \affiliation{Materials Research and Technology
  Department, Luxembourg Institute of Science and Technology (LIST),
  Avenue des Hauts-Fourneaux 5, L-4362 Esch/Alzette, Luxembourg}
\affiliation{Department of Physics and Materials Science, University
  of Luxembourg, Rue du Brill 41, L-4422 Belvaux, Luxembourg}
\author{Hugo Aramberri}\affiliation{Materials Research and Technology
  Department, Luxembourg Institute of Science and Technology (LIST),
  Avenue des Hauts-Fourneaux 5, L-4362 Esch/Alzette, Luxembourg}
\author{Tony Schenk}\affiliation{Materials Research and Technology
  Department, Luxembourg Institute of Science and Technology (LIST),
  Avenue des Hauts-Fourneaux 5, L-4362 Esch/Alzette, Luxembourg}
\author{Jorge \'I\~niguez}\affiliation{Materials Research and Technology
  Department, Luxembourg Institute of Science and Technology (LIST),
  Avenue des Hauts-Fourneaux 5, L-4362 Esch/Alzette, Luxembourg}
\affiliation{Department of Physics and Materials Science, University
  of Luxembourg, Rue du Brill 41, L-4422 Belvaux, Luxembourg}

\begin{abstract}
Films of all-important compound hafnia (HfO$_{2}$) can be prepared in
an orthorhombic ferroelectric (FE) state that is ideal for
applications, e.g. in memories or negative-capacitance field-effect
transistors. The origin of this FE state remains a mystery, though, as
none of the proposed mechanisms for its stabilization -- from surface
and size effects to formation kinetics -- is fully
convincing. Interestingly, it is known that doping HfO$_{2}$ with
various cations favors the occurrence of the FE polymorph; however,
existing first-principles works suggest that doping by itself is not
sufficient to stabilize the polar phase over the usual non-polar
monoclinic ground state. Here we use first-principles methods to
reexamine this question. We consider two representative isovalent
substitutional dopants, Si and Zr, and study their preferred
arrangement within the HfO$_{2}$ lattice. Our results reveal that
small atoms like Si can adopt very stable configurations (forming
layers within specific crystallographic planes) in the FE orthorhombic
phase of HfO$_{2}$, but comparatively less so in the non-polar
monoclinic one. Further, we find that, at low concentrations, such a
dopant ordering yields a FE ground state, the usual paraelectric phase
becoming a higher-energy metastable polymorph. We discuss the
implications of our findings, which constitute a definite step forward
towards understanding ferroelectricity in HfO$_{2}$.
\end{abstract}

\maketitle

Ever since it was shown that HfO$_{2}$ films can be prepared in a
ferroelectric (FE) phase \cite{boscke11,muller12}, much effort has
focused on elucidating the origin of such a surprising state, never
observed in bulk form. Ferroelectricity in hafnia generally becomes
more robust as the size of the grains or crystallites gets smaller
\cite{yurchuk13}, a feature that is just opposed to what is typical in
traditional ferroelectrics (e.g., perovskite oxides) and seems to
suggest that surface effects play a role in the stabilization of the
polar polymorph \cite{materlik15}. However, evidence for this
explanation is not conclusive and, more recently, other possible factors,
ranging from kinetics of formation of the ferroelectric (FE) phase to
the role of phase boundaries and local strains, have been discussed
\cite{park17,kunneth17,shimizu18,grimley18,park19,liu19}.

Interestingly, it is experimentally known that substitutional cation
dopants (e.g., Si \cite{boscke11}, Zr \cite{muller11b}, , Y
\cite{muller11a}, Al \cite{mueller12}, La \cite{chernikova16}) greatly
facilitate the formation of FE hafnia \cite{park15}. This suggests
that, for a suitably chosen dopant {\sl A}, the mixture Hf$_{1-x}${\sl
  A}$_{x}$O$_{2}$ will undergo a paraelectric (PE) to FE transition as
$x$ increases, thus displaying a {\em morphotropic phase boundary}
like many perovskite oxides do. (For example,
Sr$_{1-x}$Ba$_{x}$TiO$_{3}$ undergoes such a transformation for
increasing Ba content \cite{lemanov96}.)  First-principles methods
based on density functional theory (DFT) are ideally suited to reveal
such morphotropic transitions; however, as far as we know, the results
for doped hafnia have been negative so far: all DFT studies predict
that doping by itself is not sufficient to stabilize the FE polymorph
over the PE ground state of the compound
\cite{batra17,kunneth18,materlik17,materlik18,falkowski18,dogan19}.

The present work originated from our attempts at using doping to
improve the electromechanical responses of FE hafnia. In our
simulations with representative tetravalent dopants, it quickly became
clear that (some) doping atoms have a (very strong) preference to
adopt specific spatial arrangements. An important conclusion follows:
It is not appropriate to assume -- as implicitly done in the all cited
DFT works except for Ref.~\cite{falkowski18} -- that the dopants
locate randomly in the HfO$_{2}$ lattice. We thus focus on this issue
and address the following questions: What is the preferred spatial
configuration of substitutional cation dopants in HfO$_{2}$? Can
dopant ordering result in a stronger stabilization of the FE phase
over the PE polymorphs?

For simplicity here we consider doping with two tetravalent cations,
Si and Zr, that have been extensively studied experimentally. We
assume that the isovalent replacement of Hf by Si or Zr occurs without
any accompanying defect or charged state. Also, we focus on the lowest
energy and most common FE phase of hafnia, which has orthorhombic
symmetry $Pca2_{1}$ \cite{huan14,reyes-lillo14,barabash17} and we
denote ``FE-o'' in the following; thus, we do not consider here other
FE polymorphs recently reported \cite{barabash17,wei18}. HfO$_{2}$ has
many PE phases, but here we consider only two: the common monoclinic
phase, with $P2_{1}/c$ symmetry and denoted ``PE-m'' in the following,
which is stable at ambient conditions and constitutes the ground state
of the pure compound \cite{barabash17}; the PE tetragonal polymorph
with space group $P4_{2}/nmc$ and denoted ``PE-t'' in the following,
which has been discussed as a bridge state leading to the
stabilization of the FE-o structure \cite{huan14,shimizu18} and whose
relevance in this work will be made clear below.

\begin{figure}[t!]
\includegraphics*[width=\linewidth]{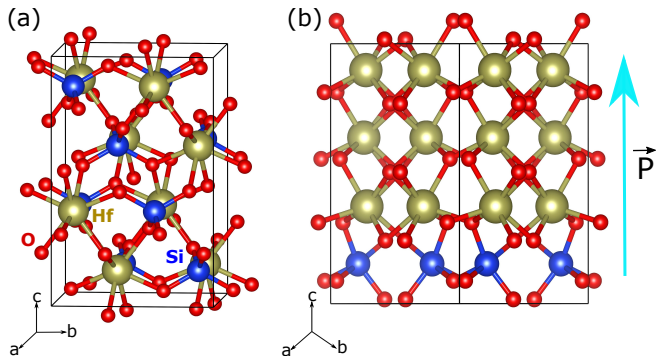}
\caption{Representative low-energy structures of the
  Hf$_{1-x}$Si$_{x}$O$_{2}$ compound, obtained from structural
  relaxations of the 48-atom supercell mentioned in the
  text. Panel~(a) corresponds to the most stable atomic ordering
  obtained for the PE-t polymorph at $x=0.5$; the Hf and Si atoms are
  intercalated. Panel~(b) corresponds to the most stable atomic
  ordering obtained for the FE-o polymorph at $x=0.25$; the Si dopants
  form a layer perpendicular to the $c$ crystallographic axis, which
  coincides with the direction of the FE polarization (marked with an
  arrow).}
\label{fig:structure}
\end{figure}

Most of our calculations are done in a 48-atom supercell containing 16
Hf$_{1-x}${\sl A}$_{x}$O$_{2}$ formula units (see
Fig.~\ref{fig:structure}). This supercell allows us to consider
composition steps $\Delta x$ of 0.0625 (6.25~\%), and we study mixings
up to $x = 0.5$ (50~\%). Whenever we have more than one dopant in the
supercell, we study a representative number of spatial arrangements,
including distinct limit cases: isolated dopants, clustering forming
quasi-spherical aggregates, dopants forming layers in different
cyrstallographic planes, dopants intercalated with the Hf atoms,
etc. All in all, we study 4 different orders for 12.5~\% doping, 12
for 25~\%, 17 for 31.25~\%, 16 for 37.5~\%, 12 for 43.75~\%, and 12
for 50~\%. For the DFT simulations we use standard methods implemented
in the VASP package \cite{kresse96,kresse99}. See Suppl. Mats. for
more details on our simulations.

We focus on the calculation of formation energies of the different
polymorphs as a function of composition. For a concentration $x$ of
dopant {\sl A}, this quantity is defined as
\begin{equation}
  E_{\rm for}(x) = E(x) - (1-x)E_{\rm HfO_{2}} - xE_{\rm {\sl A}O_{2}}
  \; .
\end{equation}
Here, $E(x)$ is the energy of the Hf$_{1-x}${\sl A}$_{x}$O$_{2}$
compound as computed for a particular polymorph and arrangement of the
{\sl A} dopants; further, $E_{\rm HfO_{2}}$ and $E_{\rm {\sl A}O_{2}}$
are the ground-state energies of the pure HfO$_{2}$ and {\sl A}O$_{2}$
materials, respectively. The PE-m phase is taken to be the ground
state of HfO$_{2}$ \cite{barabash17} and ZrO$_{2}$
\cite{reyes-lillo14}. For SiO$_{2}$ we use the $I\bar{4}2d$ structure
reported in Ref.~\cite{coh08}.

\begin{figure*}[htb]%
\includegraphics*[width=\textwidth]{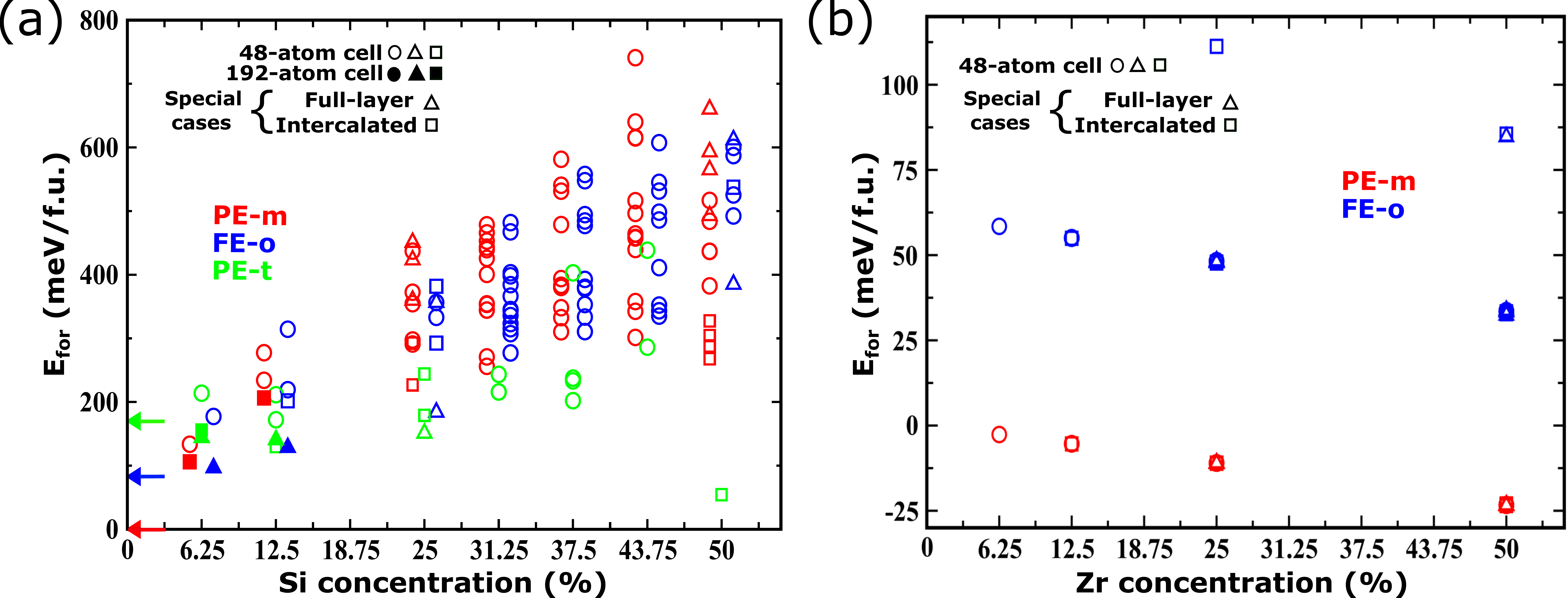}
\caption{Formation energies $E_{\rm for}$ corresponding to various
  polymorphs (marked with different colors), dopant compositions and
  dopant arrangements. Energies are given in meV per
  HfO$_{2}$-equivalent formula unit (f.u.); note that the energy scale
  is not the same in both panels. In panel~(a), at each composition
  considered, the results for the PE-m and FE-o polymorphs are
  slightly shifted horizontally, for visibility. Squares are used for
  arrangements where the Hf and dopant atoms are perfectly
  intercalated, while triangles correspond to configurations
  presenting perfect dopant layers; circles are used for any other
  dopant order, including defective layering or intercalation. The
  colored arrows placed at $x=0$ in panel~(a) mark the energies of the
  PE-m (red), PE-t (green), and FE-o (blue) polymorphs as computed for
  pure HfO$_{2}$, taking the PE-m solution as the zero of energy.}
\label{fig:energy}
\end{figure*}

Let us discuss our results following the progress of our simulations
and discoveries. Our initial calculations focused on comparing the
PE-m and FE-o polymorphs, using the mentioned 48-atom supercell. The
corresponding results are shown in Fig.~\ref{fig:energy} as open red
(PE-m) and open blue (FE-o) symbols. We find a marked difference
between the behavior of the two dopants. For Si
(Fig.~\ref{fig:energy}(a)) the energy differences between different
dopant arrangements are massive, with gaps of as much as 400~meV {\sl
  per} formula unit (f.u.)  separating the most and least stable
configurations. In contrast, the Zr dopants (Fig.~\ref{fig:energy}(b))
present a much weaker tendency towards ordering, with many low-laying
dopant orderings yielding energies within a window of 1~meV/f.u.

It is also worth noting that, for Si doping, formation energies are
always positive, implying that Hf$_{1-x}$Si$_{x}$O$_{2}$ mixtures are
metastable. In contrast, the PE-m phase of Hf$_{1-x}$Zr$_{x}$O$_{2}$
presents negative formation energies, indicating that there is a
thermodynamic drive towards forming such solid solutions. More
interestingly, for Si doping the PE-m and FE-o structures occupy the
same energy range and clearly compete; in contrast, the PE-m state is
clearly dominant for all considered Zr compositions. In the following
we focus on the Si doping, which is clearly more intriguing.

We find that, in some cases, the relaxation of the Si-doped PE-m and
FE-o structures yields a different solution, namely, the PE-t
polymorph represented with green open symbols in
Fig.~\ref{fig:energy}(a). The PE-t state can be very stable; in
particular, the Si-doped structure with the lowest formation energy,
which occurs for a Si concentration of 50~\% and is shown in
Fig.~\ref{fig:structure}(a), has this symmetry. Having found that the
PE-t state becomes relevant upon doping, we run for this polymorph all
the dopant arrangements previously considered for PE-m and FE-o at
6.25~\% and 12.5~\% concentrations. The results, represented by open
green symbols in Fig.~\ref{fig:energy}(a), show that the PE-t state
becomes dominant as the amount of Si grows. (A similar stabilization
of the PE-t has already been reported in Ref.~\cite{falkowski18}.)
Even more interestingly, our results also show that the FE-o polymorph
becomes more stable than the PE-m state for Si concentrations of
12.5~\% to 25~\%.

\begin{figure}[t]%
\includegraphics*[width=\linewidth]{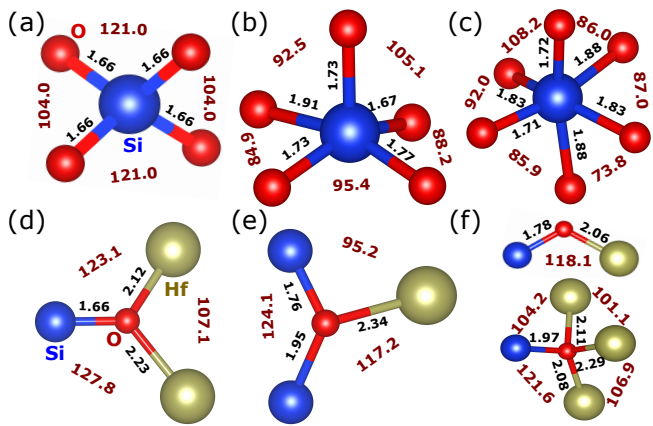}
\caption{Representative local environments of Hf and O atoms, as
  obtained from our simulations (see text). We indicate bond distances
  (in \AA) and bond angles (in degrees).}
\label{fig:local}
\end{figure}

By inspecting the structures of the lowest- and highest-energy dopant
arrangements, we can identify the features resulting in more stable
configurations. Our findings are summarized in Fig.~\ref{fig:local},
where Si--O and Hf--O pairs separated by less than 2.4~\AA\ are
displayed as forming a chemical bond. Let us start by noting that Si
is considerably smaller than Hf. More specifically, the tabulated
covalent radii of Hf and Si are 1.87~\AA\ and 1.11~\AA, respectively
\cite{cordero08}; as for the ionic radii of Hf$^{4+}$ and Si$^{4+}$,
we have 0.7~\AA\ and 0.4~\AA, respectively \cite{shannon76}. This size
difference suggests that, as compared with Hf, the Si dopants will
prefer relatively small oxygen coordination numbers, as we indeed
observe in our results. More specifically, we find that some of the
most stable structures display SiO$_{4}$ groups forming nearly regular
tetrahedra, as shown in Fig.~\ref{fig:local}(a). In particular, the
SiO$_{4}$ coordination is typical of the doped tetragonal polymorph,
and it characterizes the lowest-energy PE-t solutions, including the
structure shown in Fig.~\ref{fig:structure}(a). A second usual
coordination involves SiO$_{5}$ groups forming a (distorted)
square-base pyramid, as shown in Fig.~\ref{fig:local}(b). All Si
dopants present this kind of environment in the lowest-lying PE-m and
FE-o states, as is the case of the polar structure in
Fig.~\ref{fig:structure}(b). Finally, we also find higher Si
coordinations, as e.g. the quasi-octahedral SiO$_{6}$ groups depicted
in Fig.~\ref{fig:local}(c), that are typical of high-lying PE-m and
FE-o configurations.

Related trends can be identified by paying attention to the
environment of the oxygens: Without exception, in all our lowest-lying
structures each oxygen is bound to one Si and two Hf atoms, as
depicted in Fig.~\ref{fig:local}(d); in contrast, structures
displaying coordination complexes like that of Fig.~\ref{fig:local}(e)
lie at higher energies, and features as those in
Fig.~\ref{fig:local}(f) are typical of even less stable cases. Hence,
our results suggest that the lowest-energy states are those whose
chemical-bond topology allows the dopants to form the most stable
bonding complexes, namely, the ones in Figs.~\ref{fig:local}(a),
\ref{fig:local}(b) and \ref{fig:local}(d).

Interestingly, a group of low-energy states present a long-range order
whereby the Si dopants are intercalated with the Hf atoms, as in the
case of Fig.~\ref{fig:structure}(a); this order, represented by
squares in Fig.~\ref{fig:energy}(a), is typical of the lowest-lying
PE-m structures and some of the most stable PE-t states. In contrast,
the remaining low-energy structures are characterized by a layering of
the Si dopants; such layered structures, represented by triangles in
Fig.~\ref{fig:energy}(a) and illustrated in
Fig.~\ref{fig:structure}(b), are typical among the lowest-lying FE-o
solutions and some of the most stable PE-t states. Nevertheless, as
the data in Fig.~\ref{fig:energy}(a) show, intercalation or layering
alone do not guarantee that the doped structure will have a low
formation energy. Indeed, specific details of the ordering are
critical for a low-energy state to occur: most importantly, in the
most stable Si-doped FE-o solutions, the dopant layers are
perpendicular to the polar $c$ axis (Fig.~\ref{fig:structure}(b)),
while in the most stable layered PE-t structures the dopants occupy
planes perpendicular to the $a$ crystallographic axis.

\begin{figure}[t]%
\includegraphics*[width=\linewidth]{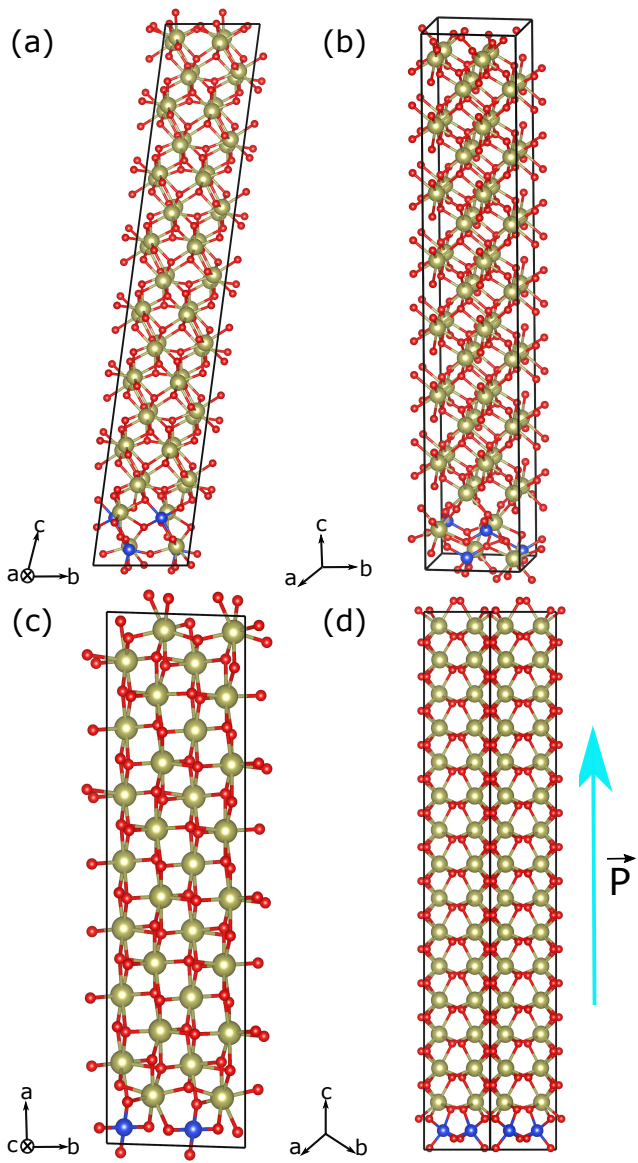}
\caption{192-atom supercells used to investigate, in the limit of very
  low dopant concentrations, the relative stability of the most
  favorable dopant arrangements identified in this work. Panel~(a):
  PE-m polymorph with a region of intercalated Hf/Si atoms; (b): PE-t
  polymorph with a region of intercalated Hf/Si atoms; (c): PE-t
  polymorph with an $a$-oriented dopant layer; (d): FE-o polymorph
  with a $c$-oriented dopant layer.}
\label{fig:supercells}
\end{figure}

One may wonder whether our 48-atom supercell is realistic to
investigate the relative stability of these three structures in the
limit of small dopings; for example, for a doping concentration of
6.25~\%, the size and shape of this supercell are incompatible with
having a complete dopant layer, and only isolated dopants can be
studied. Can the peculiar orders observed at higher concentrations,
which favor the FE-o and PE-t states, result in the stabilization of
such structures for lower doping levels? To test this, we run
simulations in elongated 192-atom supercells as those shown in
Fig.~\ref{fig:supercells}, considering the most energetically
favorable dopant arrangements previously identified: a localized
region with intercalated Hf/Si atoms for PE-m
(Fig.~\ref{fig:supercells}(a)) and PE-t (Fig.~\ref{fig:supercells}(b))
polymorphs, and suitably-oriented layers of Si dopants for PE-t
(Fig.~\ref{fig:supercells}(c)) and FE-o (Fig.~\ref{fig:supercells}(d))
structures. The corresponding formation energies are shown as filled
symbols in Fig.~\ref{fig:energy}(a). These results indicate that the
FE-o state with full Si layers constitutes the lowest-energy solution
at 6.25~\%, thus predicting that the FE-o phase is the
thermodynamically stable ground state of HfO$_{2}$ upon moderate Si
doping! Then, as the Si concentration increases to 12.5~\%, the FE-o
(layered) and PE-t (intercalated) solutions become essentially
degenerate, and for higher dopant contents the PE-t state dominates.

Let us emphasize our results for the 6.25~\% concentration, as they
capture very well the main message of this work. In this case, the
open circles in Fig.~\ref{fig:energy}(a) pertain to isolated-impurity
calculations as those typically reported in previous DFT studies
\cite{batra17,kunneth18,materlik17,materlik18,dogan19}. The
corresponding formation energies largely reflect the energy
differences between the pure HfO$_{2}$ polymorphs, marked by the
colored arrows at $x=0$ in the figure; hence, the PE-m state
prevails. However, when the dopants are allowed to adopt their
lowest-energy configuration (filled symbols in the figure), strong
energy reductions are obtained for the PE-t and FE-o cases, while the
energy gain is comparatively small for the PE-m polymorph. This
divergent behavior is the key to stabilize the FE-o phase over the
PE-m state.

For the FE-o phase of pure HfO$_{2}$, we calculate a polarization of
55~$\mu$C/cm$^{2}$, in agreement with previous literature
\cite{huan14,clima14,reyes-lillo14}. For the doped structure with a
25~\% Si concentration (Fig.~\ref{fig:structure}(b)), we obtain
44~$\mu$C/cm$^{2}$. Further, for the 6.25~\% structure in
Fig.~\ref{fig:supercells}(d) we obtain 53~$\mu$C/cm$^{2}$. Hence, we
predict that the Si dopants stabilize the FE-o phase without harming
its ferroelectric polarization.

Our findings seem to be mainly related with the size and preferred
oxygen coordination of the relatively-small Si dopants, and the
ability of the FE-o and PE-t polymorphs to better accommodate them. If
this interpretation is correct, similar behaviors can be expected for
other small dopants. Indeed, we have preliminary first-principles
evidence that Ge too stabilizes the FE-o polymorph over the PE-m phase
when a full $c$-oriented layer is formed (i.e., when we have a
situation analogous to that of
Fig.~\ref{fig:structure}(b)). Additionally, our results for Zr suggest
that ordering is not expected to occur for relatively large
dopants. (Our preliminary first-principles results indicate that the
behavior of Ti, Sn and Pb dopants is similar.) Further, the observed
prevalence of the PE-m state upon doping (Fig.~\ref{fig:energy}(b))
suggests that the connection between big dopants and the stabilization
of polar HfO$_{2}$ is, at best, indirect.

Our theoretical findings have practical implications worth
discussing. Most importantly, they suggest that, by depositing thin
SiO$_{2}$ layers during the growth of hafnia films (so that the silica
content adds up to about 6--12~\% of the material), one should be able
to reliably obtain samples in the FE-o phase. The need for a
``wake-up'' step to observe the FE behavior (as is typical in hafnia
thin films \cite{zhou13,schenk14}) would be much reduced in such
samples, if present at all, as the FE-o phase is their thermodynamic
ground state. The predicted most stable geometry, with polarization
lying along the growth (out-of-plane) direction, is all but ideal to
maximize the remnant polarization of the films.

This proposed preparation strategy could be realized by employing
techniques that allow great control of the epitaxial growth, like
e.g. pulsed laser deposition. Nevertheless, it is important to note
that our ideal scenario is strongly reminiscent of how most FE hafnia
films are actually grown, via atomic layer deposition (ALD) where the
dopant ratio is achieved by performing a dopant oxide ALD-cycle after
a certain number of HfO$_{2}$ cycles \cite{park15}. ALD-grown films
are then subject to a thermal treatment to induce crystallization;
while dopants may diffuse at this step, the resulting samples still
present a modulation in dopant concentration along the growth
direction, thus displaying diffuse dopant layers
\cite{lomenzo15,richter17}. Hence, when the ALD samples are subject to
the wake-up treatment -- i.e., application of alternate electric
fields, typically along the out-of-plane direction --, they are
(according to our results) suitably {\em pre-conditioned} to yield the
FE-o phase. We thus believe the present findings are consistent with
the usual experimental route to obtain FE hafnia, and partly explain
why it works. They also indicate that controlling the layering of
small dopants may be a key to produce better samples.

Having said this, we should bear in mind that our study is limited to
ideal phases of doped HfO$_{2}$, the connection with experiment being
far from perfect. ALD films are randomly oriented \cite{schenk19},
which implies that, in principle, a good alignment between the dopant
layers and specific crystallographic axes (e.g., the polar axis of the
FE-o phase) will be possible only in a fraction of the
grains. Similarly, the wake-up process is a complex one, known to
involve many effects: from transformation of non-polar regions into
polar ones \cite{grimley16} and re-orientation of ferroelastic domains
\cite{shimizu18} to movement of defects \cite{zhou13} and modification
of internal bias fields \cite{schenk15}. Hence, it would be naive to
take our predictions as a fail-safe strategy to obtain perfect samples
of FE hafnia. Yet, they do provide us with a definite motivation to
explore directions focused on dopant ordering.

We conclude by noting that, HfO$_{2}$ being such a polymorphic
material, the present study could be extended by considering dopant
ordering in other low-lying metastable phases. Further, one could also
use advanced DFT methods for structure discovery
\cite{oganov-book2010,wang12} to identify even more stable dopant
arrangements. Such investigations might impact the predicted relative
stability of doped-HfO$_{2}$ polymorphs, and are worth tackling in the
future. Nevertheless, notwithstanding future extensions and
improvements, the basic conclusion of the present work -- namely, that
the spatial arrangement of small dopants in HfO$_{2}$ is critically
important -- is clear and robust. Further, its most obvious
consequence -- that, when accounting for the possible spatial
arrangements of the dopants, the formation energies of the dominant
polar and non-polar polymorphs fall within the same range -- also
seems robust. Hence, the present results should definitely change the
way we think of dopants in HfO$_{2}$ and how they help stabilize its
polar phase.

Work funded by the Luxembourg National Research Fund (FNR) through
Grants PRIDE/15/10935404 ``MASSENA'' (S.D. and J.\'I.),
INTER/ANR/16/11562984 ``EXPAND'' (H.A. and J.\'I.). T.S. acknowledges
the finacial support of LIST through project SF\_MRT\_CSDFO.

\end{document}